\documentclass[aps,pra, twocolumn,floats,superscriptaddress]{revtex4-1}
\usepackage{epsfig,color}
\usepackage{bm}
\usepackage{amsmath}
\usepackage{subfigure}
\usepackage{hyperref}

\begin{document}

\author{Lei Wang}
\affiliation{Theoretische Physik, ETH Zurich, 8093 Zurich, Switzerland}
\author{Troels F. R{\o}nnow}
\affiliation{Theoretische Physik, ETH Zurich, 8093 Zurich, Switzerland}
\author{Sergio Boixo}
\affiliation{Information Sciences Institute and Department of Electrical Engineering, University of Southern California, Los Angeles, CA 90089, USA}
\author{Sergei V. Isakov}
\affiliation{Theoretische Physik, ETH Zurich, 8093 Zurich, Switzerland}
\author{Zhihui Wang}
\affiliation{Department of Chemistry and Center for Quantum Information Science \& Technology,  University of Southern California, Los Angeles, California 90089, USA}
\author{David Wecker}
\affiliation{Quantum Architectures and Computation Group, Microsoft Research, Redmond, WA 98052, USA}
\author{Daniel A. Lidar}
\affiliation{
Departments of Electrical Engineering, Chemistry and Physics, and Center for Quantum Information Science \& Technology, University of Southern California, Los Angeles, California 90089, USA
}
\author{John M. Martinis}
\affiliation{Department of Physics, University of California, Santa Barbara, CA 93106-9530, USA}
\author{Matthias Troyer}
\affiliation{Theoretische Physik, ETH Zurich, 8093 Zurich, Switzerland}

\title{Comment on: ``Classical signature of quantum annealing"}

\begin{abstract}
In a recent preprint (arXiv:1305.4904) entitled ``Classical signature of quantum annealing" Smolin and Smith point out that a bimodal distribution presented in (arXiv:1304.4595) for the success probability in the D-Wave device does not in itself provide sufficient evidence for quantum annealing, by presenting a classical model that also exhibits bimodality. Here we analyze their model and in addition present a similar model derived from the semi-classical limit of quantum spin dynamics, which also exhibits a bimodal distribution. We find that in both cases the correlations between the success probabilities of these classical models and the D-Wave device are weak compared to the correlations between a simulated quantum annealer and the D-Wave device. Indeed, the evidence for quantum annealing presented in arXiv:1304.4595 is not limited to the bimodality, but relies in addition on the success probability correlations between the D-Wave device and the simulated quantum annealer. The Smolin-Smith model and our semi-classical spin model both fail this correlation test.
\end{abstract}

\maketitle

Whether the devices built by D-Wave \cite{Harris2010} exhibit quantum effects on a large scale is a subject that has attracted significant attention recently. Quantum tunneling involving $8$ superconducting flux qubits was recently demonstrated in such a device \cite{Johnson2011}. To test whether quantum effects persist on a scale of more than $100$ qubits we used a D-Wave One device with $108$ functional qubits to run millions of experiments in which the device attempted to find the ground state of $1000$ randomly selected Ising spin glass instances \cite{QA}. For different values of the number of spins $N\in [8,108]$ we ran each instance $1000$ times and created a histogram of success probabilities. We found that this histogram is bimodal once  $N$ becomes large, with a higher number of ``easy" and ``hard" instances (high and low success probabilities, respectively) than intermediate instances. We also performed quantum Monte Carlo (simulated quantum annealing, SQA) simulations for the same Ising model instances and observed a very similar bimodal histogram for each $N$. In contrast, classical simulated annealing (SA) calculations for the same problem instances yielded a unimodal histogram with a large concentration of instances at intermediate success probabilities.

Very recently a preprint by Smolin and Smith appeared, questioning whether such histograms are sufficient to establish that the D-Wave device exhibits quantum effects \cite{Smolin}. To this end they presented a classical ``compass needle" spin model which also exhibits a bimodal histogram for random instances satisfying the constraints of the D-Wave One hardware graph. We concur that the mere difference between a bimodal and unimodal distribution is insufficient, and will argue here that the Smolin-Smith model in fact supports our central conclusion that the D-Wave device is a quantum annealer. 

To this end we note that bimodality versus unimodality alone was indeed \emph{not solely} the basis for our conclusion that the D-Wave device performs quantum annealing with more than $100$ qubits. A key additional piece of evidence we presented in \cite{QA}, which was unaccounted for by Smolin and Smith, are the correlations between the success probabilities of instances for the D-Wave device and the simulated quantum annealer. As discussed in detail in our work \cite{QA} the ``easy" and ``hard" peaks in both cases contain similar numbers of instances and their hardness correlates, i.e., \emph{the same subset of instances is hard or easy for both the device and the simulated quantum annealer}. 

In the classical O(2) spin-model presented in \cite{Smolin} such correlations are absent, which can already be deduced from the fact that the two peaks in the bimodal distribution presented there have a very different weight distribution compared to both our D-Wave device measurements and numerical SQA results. There are many more ``hard'' instances in the model of Ref. \cite{Smolin}.

Thus the burden of proof on classical models attempting to replicate the statistics of the D-Wave One measurement outcomes is much heavier than a demonstration of bimodality, which is merely a necessary condition (not satisfied by SA, as shown in \cite{QA}): the correct success probability correlations should also be a feature of any such model.

To illustrate this aspect we investigated a model related to that of Smolin and Smith, based on the semi-classical limit of quantum spin dynamics. This model more
closely resembles the classical dynamics of the system than the O(2) model of \cite{Smolin}, and replaces the quantum spins by O(3) classical unit vectors $\vec{M}_i$ propagated via the equations of motion
\begin{eqnarray}
  \frac{\partial \vec{M}_i}{\partial t}  = \vec{H}_i(t)
  \times \vec{M}_i \ ,
\end{eqnarray}
where the time-dependent field $\vec{H}_i(t)$ acting on spin $i$ is a sum of a decaying transverse field (along the $x$ direction) and a growing coupling term (along $z$):
\begin{eqnarray}
\vec{H}_i(t) \equiv  (1-t/t_f)  h \hat{e}_x- (t/t_f) \sum_j J_{i j} M^z_j \hat{e}_z \ ,
\label{eq:O3}
\end{eqnarray}
where $t_f$ is the total evolution time. This model is equivalent to a mean-field model of the quantum spin-1/2 model where using Hartree-Fock decoupling the  wave function is a product state, and we represent each SU(2) spinor as an SO(3) vector. 

The initial condition is to have all spins aligned along the $x$ direction: $\vec{M}_i(0)  =  (-1,0,0)$. To introduce randomness we perturb this ideal initial condition and give each spin a small random kick:
\begin{equation}
  \vec{M}_i(0)  =  (- \sqrt{1 - \delta_i^2 - \eta_i^2}, \delta_i, \eta_i),
\end{equation}
where $|\delta_i|$, $|\eta_i|<0.1$. At the end of the evolution we use the  sign of the $z$-component of each spin as the value of the Ising variable in the optimization problem.

We can furthermore add a damping term by replacing
\begin{equation}
  \vec{H}_i  \rightarrow  \vec{H}_i + \alpha ( \vec{H}_i \times \vec{M}_i).
\end{equation}
The resulting equation is the Landau-Lifshitz-Gilbert equation, which is
widely used to describe micromagnetism \cite{Gilbert:04}. The additional damping term will
force the spins to approach the axis of the effective local magnetic field and  release energy. This does not change the distribution much but pushes it to the extremes (probability $0$ or $1$), thus making it more deterministic.

\begin{figure}[t]
\centering
\includegraphics[width=\columnwidth]{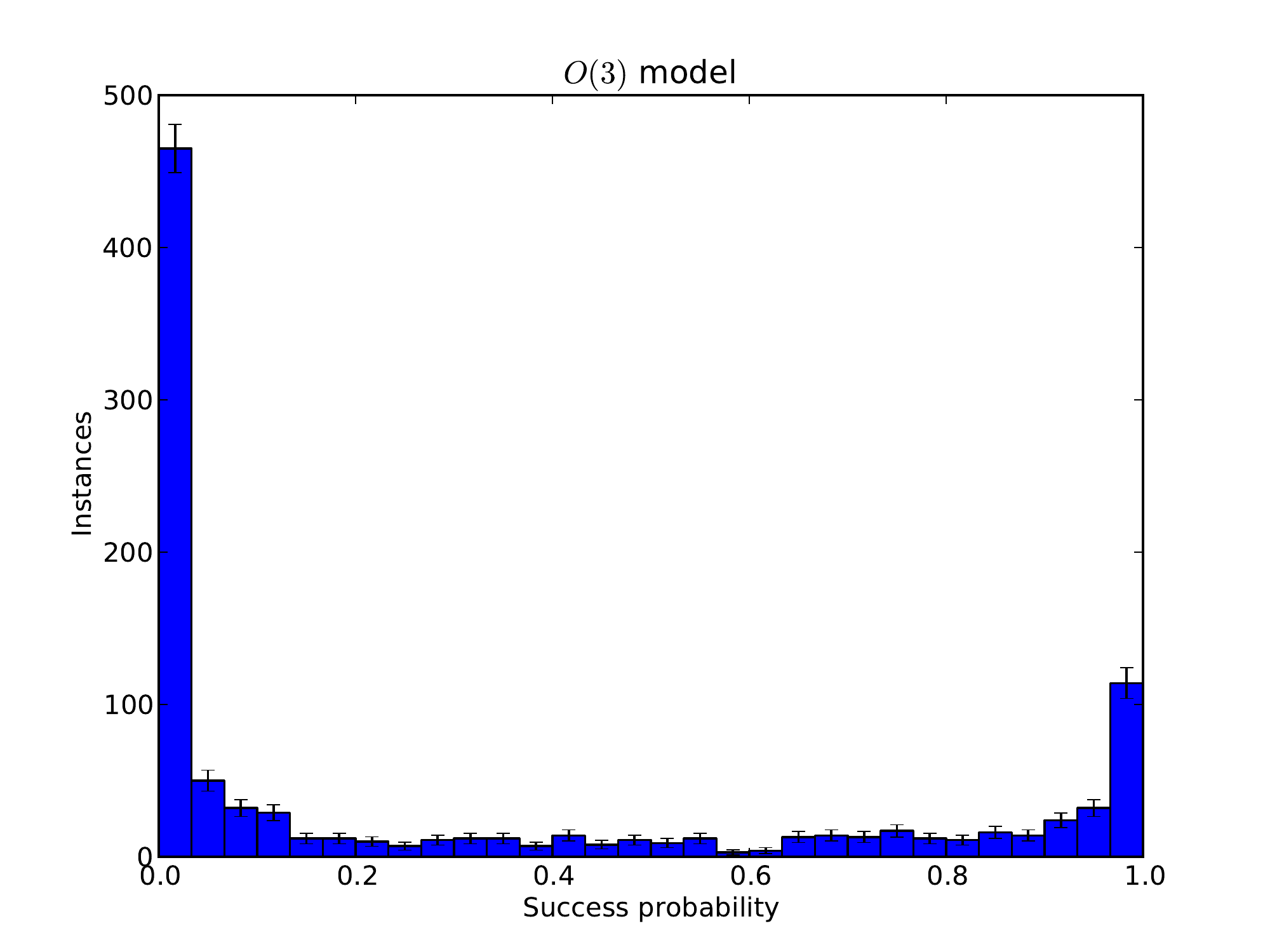}
\caption{Histogram of the success probabilities of the O(3) semi-classical annealer for $1000$ random instances.}
\label{fig:histogram}
\end{figure}

Figure~\ref{fig:histogram} shows a histogram of the success probabilities for this semi-classical O(3) annealer on $1000$ instances of a spin glass with $J_{ij}=\pm1$ on the $108$-site chimera graph of the D-Wave One device. As in \cite{Smolin} we observe a bimodal distribution for this semi-classical model, but with higher weight in the low-probability (``hard'') peak than observed in the simulated quantum annealer and the D-Wave device \cite{QA}. Compare with figure~3 in \cite{Smolin} which similarly shows a larger ``hard" peak.

\begin{figure}[b]
\centering
\includegraphics[width=\columnwidth]{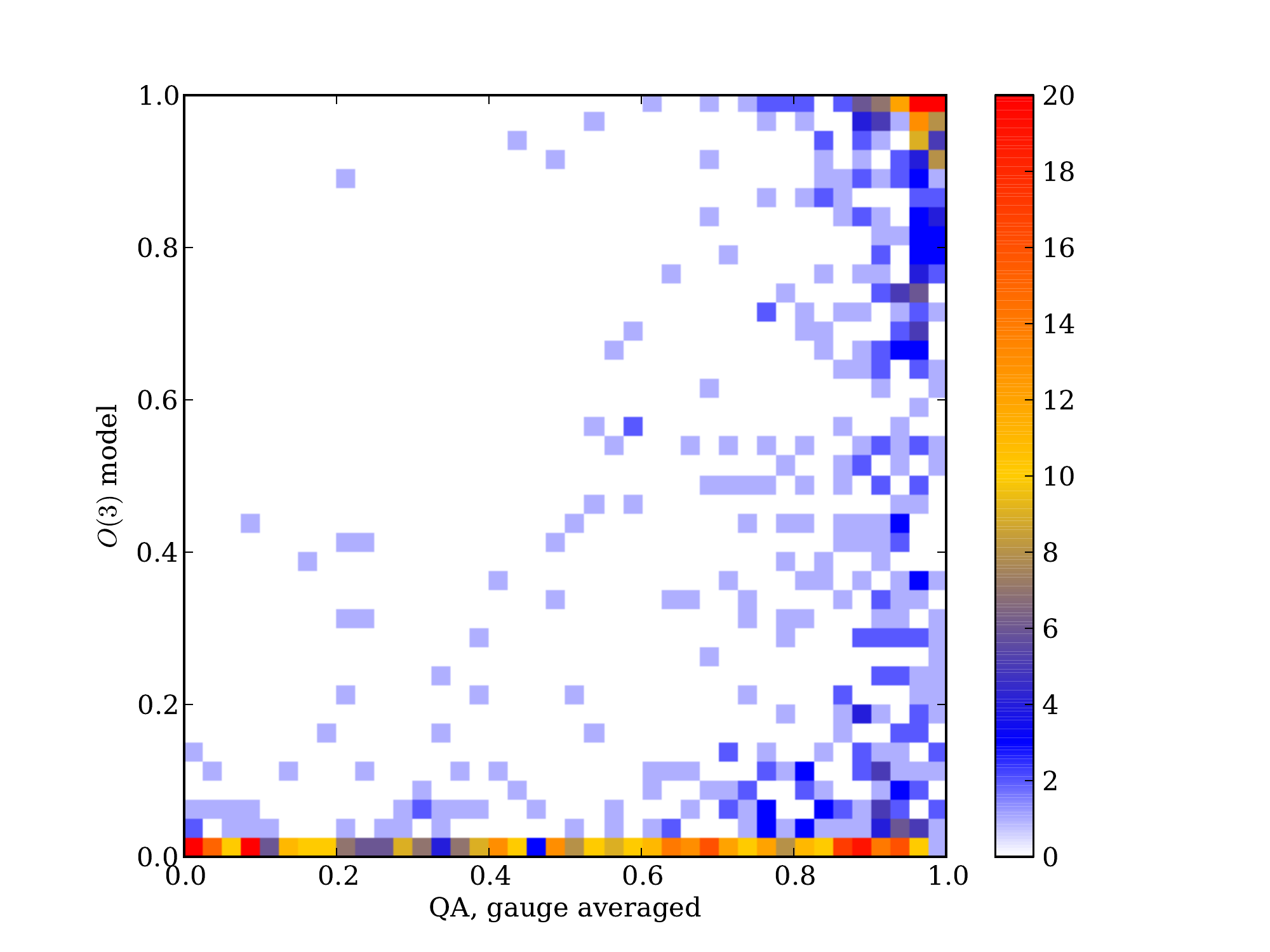}
\caption{Correlations of success probabilities between the semi-classical O(3) annealer and the D-Wave device. Note that the D-Wave machine succeeds with high probability on certain instances which the semi-classical model finds hard.}
\label{fig:correlations}
\end{figure}

The difference is more pronounced when one considers the {\em correlations} with gauge-averaged success probabilities of the same instances on the D-Wave One device, as shown in figure~5C) of \cite{QA}. That figure, which exhibits a strong positive correlation for all values of the success probability, should be compared to the correlations found between the D-Wave One device and the semi-classical annealer, shown here in figure \ref{fig:correlations}. Some instances are easy for both the semi-classical annealer and the D-Wave device (and a simulated quantum annealer); for such instances a simple semi-classical algorithm finds the ground state with high probability. However, strikingly there is no apparent correlation between the hard instances on the semi-classical annealer and the success probability on the D-Wave device. Nor does there appear to be a correlation for instances of intermediate hardness, in contrast to the correlations seen in figure~5C) of \cite{QA}.

\begin{figure}[t]
\centering
\includegraphics[width=\columnwidth]{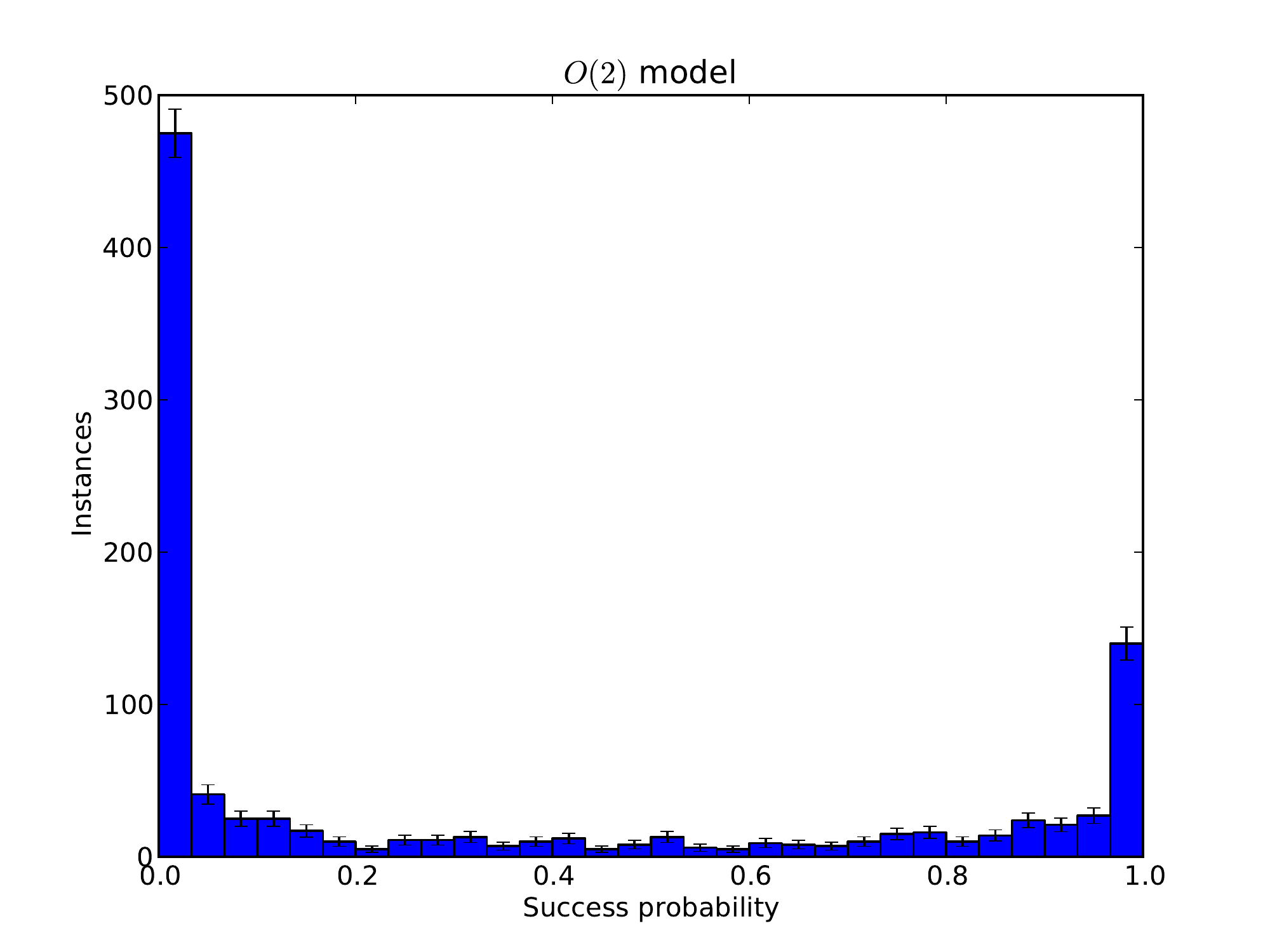}
\caption{Histogram of the success probabilities of the Smolin-Smith O(2) model \cite{Smolin}.}
\label{fig:histogramo2}
\end{figure}

\begin{figure}[tb]
\centering
\includegraphics[width=\columnwidth]{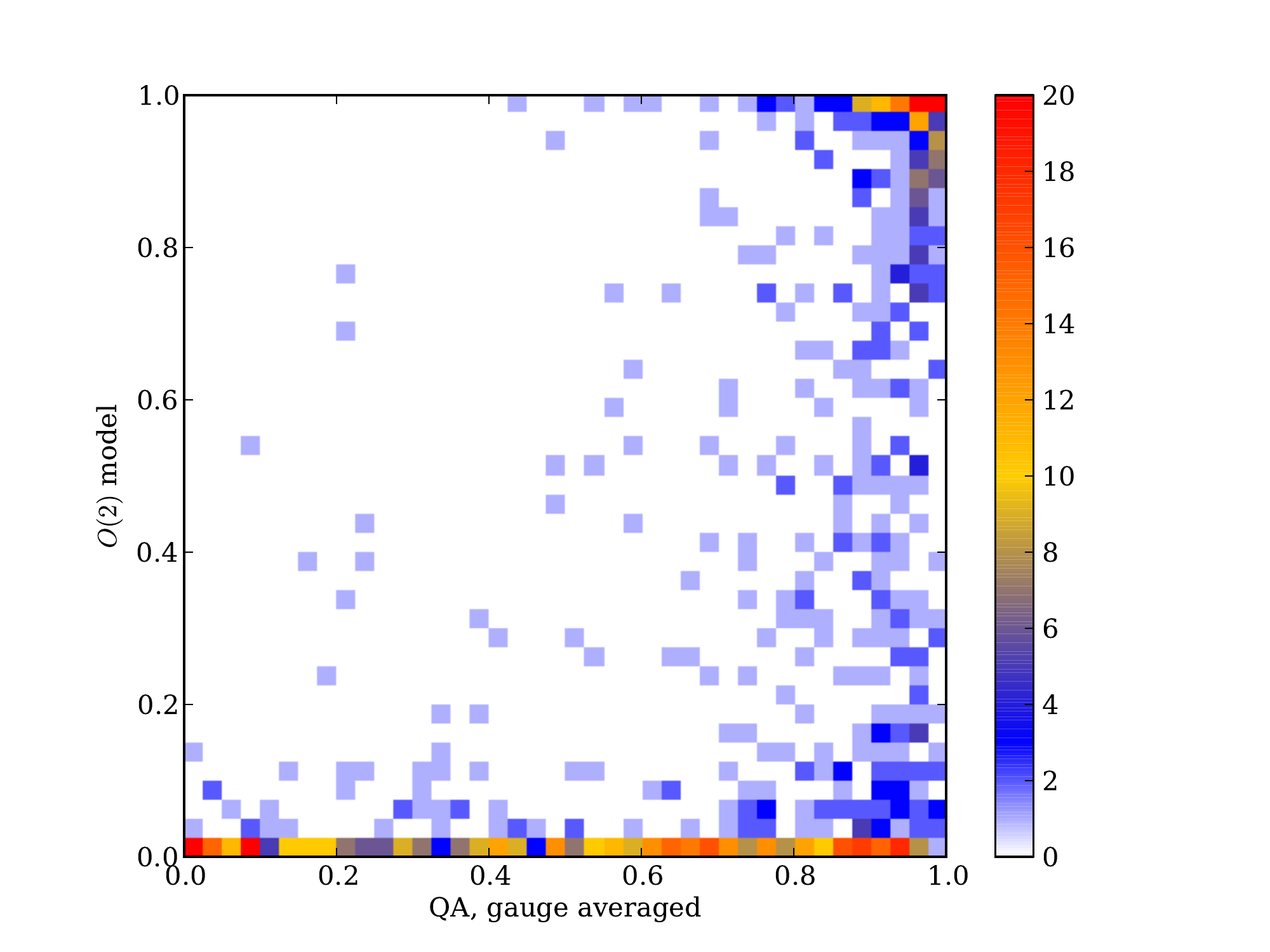}
\caption{Correlations of success probabilities between the Smolin-Smith O(2) model and the D-Wave device.}
\label{fig:correlationso2}
\end{figure}

\begin{figure}[tb]
\centering
\includegraphics[width=\columnwidth]{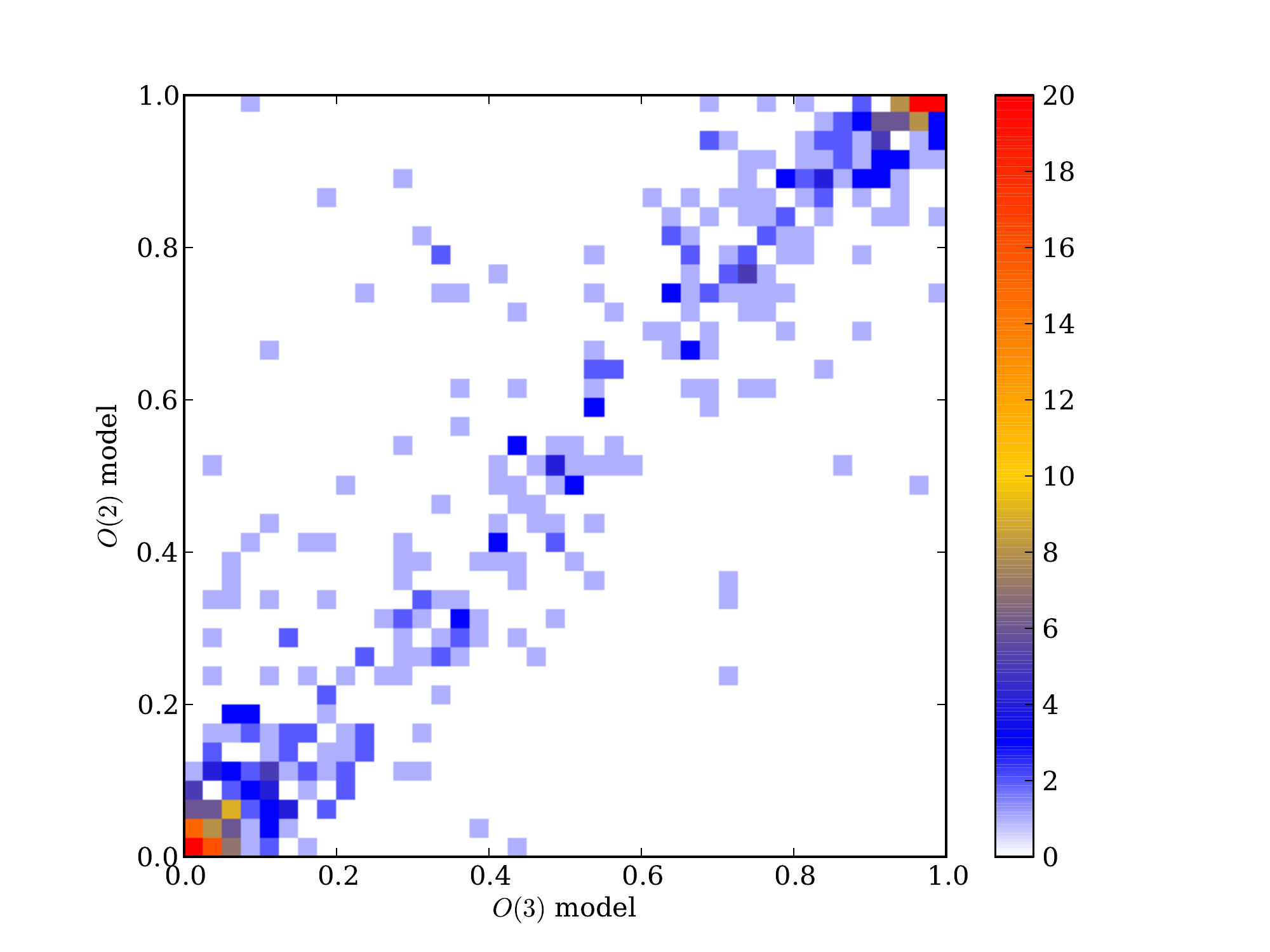}
\caption{Correlations of success probabilities between the Smolin-Smith O(2) model~\cite{Smolin} and the O(3) model Eq.(\ref{eq:O3}).}
\label{fig:O2O3}
\end{figure}

We have performed a similar analysis for the model of Ref. \cite{Smolin}, which we refer to as an O(2) model since it uses planar rotors. We show a histogram obtained by our implementation of their model in figure~\ref{fig:histogramo2} and the correlation plot in figure~\ref{fig:correlationso2}. Again the correlations are weak and this model does not explain the behavior of the D-Wave device. Adjustment of the noise levels in the O(2) and O(3) models does not improve the correlations. 
Correlating our semi-classical O(3) model with the O(2) model we find that both  these models give similar success probabilities as shown in figure \ref{fig:O2O3}. In fact, this figure resembles the correlations between the D-Wave device (after gauge averaging) and SQA, as seen in figure 5C) of \cite{QA}.

Additional evidence is presented in \cite{QA} for quantum effects beyond success probability correlations, where we showed numerically that the hard instances are characterized by small gaps in the spectrum during the annealing, while the easy instances typically have large gaps. We also showed experimentally that for those runs where the ground state is not found, easy instances tend to have excited states that are a small Hamming distance away from the ground state, while hard instances have a large Hamming distance. As explained in \cite{QA}, this agrees with instances being hard due to small gap avoided level crossings.  

To summarize, Smolin and Smith are correct to point out that it is insufficient to compare the modality of histograms in order to infer that the D-Wave device performs quantum annealing. Indeed, the claim made to this effect in \cite{QA} is based on additional evidence, in particular a strong positive correlation between the success probabilities of instances for the D-Wave device and a simulated quantum annealer, a test which the semi-classical spin models presented here and in \cite{Smolin} fail.

The question of why SQA and semi-classical spin models correlate so differently with the D-Wave device is obviously important and interesting. We note that while SQA captures decoherence in the instantaneous energy eigenbasis of the system, so that each energy eigenstate---in particular the ground state---is itself a coherent superposition of computational basis states, semi-classical spin models assume that each qubit decoheres locally, thus removing all coherence from the ground state. We conjecture that the fact that the D-Wave machine succeeds with high probability on certain instances which the semi-classical models finds hard, can be understood in terms of this difference.

\appendix
\section{Data set}
As a service to the community we provide the complete data set of 1000 instances presented in \cite{QA} for $N=108$ and the corresponding QA success probabilities as ancillary file of this e-print.

\bibliographystyle{apsrev}
\bibliography{qa}

\end{document}